\documentclass[useAMS,usenatbib,usegraphicx]{mn2e}

\usepackage{times}
\usepackage{epsfig}
\usepackage{epstopdf}
\newcommand{\kms}{\mbox{${\rm km\;s^{-1}}$}}

\newcommand{\sauron}{{\texttt {SAURON}}}

\newcommand{\HST}{{\it HST\/}}
\newcommand{\Oiii}{[{\sc O$\,$iii}]}

\newcommand{\Ha}{H$\alpha$}
\newcommand{\Hb}{H$\beta$}

\newcommand{\Nii}{[{\sc N$\,$ii}]}

\newcommand{\OiiioHb}{\Oiii/\Hb}

\newcommand{\mnras}{MNRAS}

\newcommand{\placefigNarrowBand}{
  \begin{figure}
    \begin{center}
      \includegraphics[width=\columnwidth, bb = 8 28 400 424]{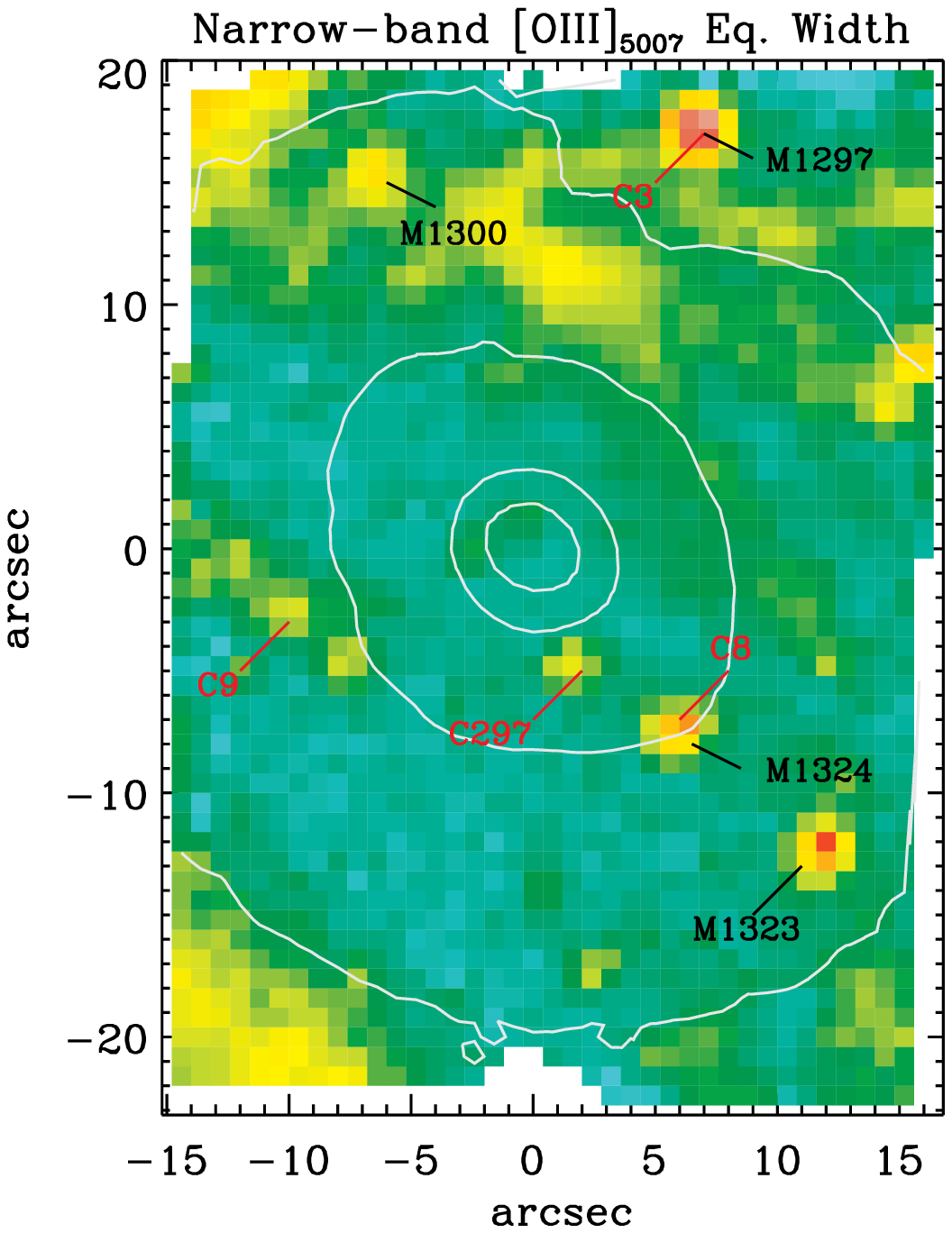}
    \end{center}
    \caption[]{Map of the equivalent width (EW) of the estimated
      \Oiii$\lambda5007$ flux from narrow-band imaging based on the
      \sauron\ data. The flux in the \Oiii\ region is measured through
      a 2.5\AA-wide spectral window centred on the expected position
      of \Oiii\ given the systemic velocity of $V_{\rm sys}=-300\,\kms$ for
      M31, whereas the flux of the stellar continuum is computed using
      the mean spectral energy density across the entire spectrum
      times the same 2.5\AA\ wavelength interval. The difference
      between these two flux measurements gives an estimate of the
      \Oiii\ flux, which once divided by the mean spectral energy
      density in the stellar continuum leads to the mapped EW values.
      The map shows both the presence of diffuse gas and of isolated
      patches of \Oiii\ emission that could originate from the
      unresolved \Oiii\ emission of PNe, more or less embedded in the
      diffuse gas component. Several of the brightest sources in this
      figures were already identified as PNe by \citet[][red
        labels]{Cia89} and \citet[][black labels]{Mer06}. In this and
      subsequent \sauron\ maps of M31, north is up an east is to the
      left.}
    \label{fig:NarrowBandmap}
  \end{figure}
}

\newcommand{\placefigAoNmap}{
  \begin{figure}
    \begin{center}
      \includegraphics[width=\columnwidth, bb = 8 28 400 424]{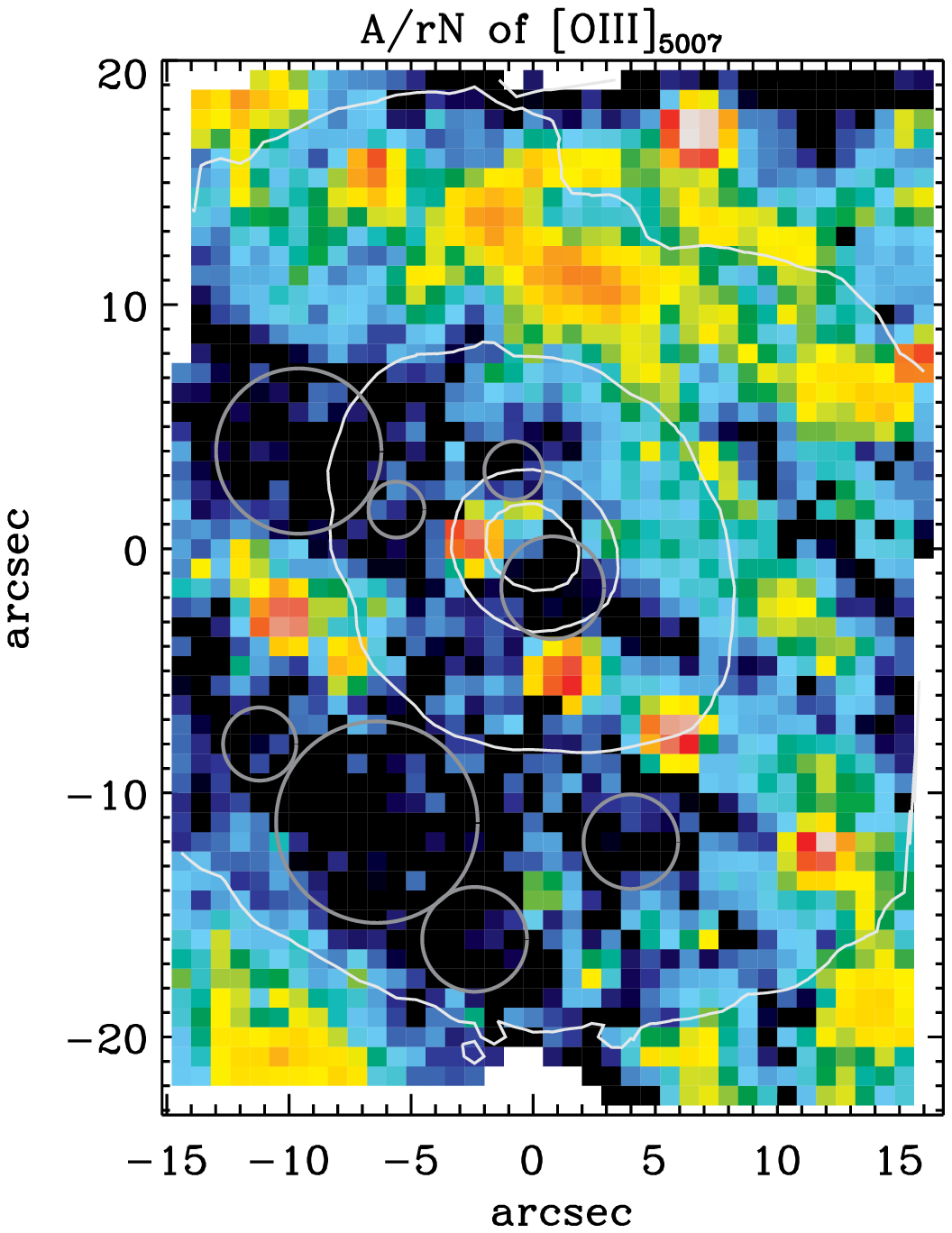}
    \end{center}
    \caption[]{Map of the values of the $A/rN$ ratio for the
      \Oiii$\lambda5007$ lines following our analysis of the
      \sauron\ spectra. The $A/rN$ values are shown in a logarithmic
      scale (from values of 1 to 64) and dark-blue bins correspond to
      regions where $A/rN < 4$ and the \Oiii\ lines are not formally
      detected. The grey circles in these regions show the aperture
      over which the \sauron\ spectra were added up to obtain
      high-quality spectra that formed the basis for our description
      of the stellar continuum in our data. Both the extent of the
      diffuse ionised-gas emission and the presence of a number of
      \Oiii\ unresolved sources consistent with PNe can be better
      appreciated here compared to Fig.~\ref{fig:NarrowBandmap}.  An
      additional known PN near the centre M31 \citep{Bac01b} is now
      also revealed.}
    \label{fig:AoNmap}
  \end{figure}
}

\newcommand{\placefigVOIIImap}{
  \begin{figure}
    \begin{center}
      \includegraphics[width=\columnwidth, bb = 8 28 400 424]{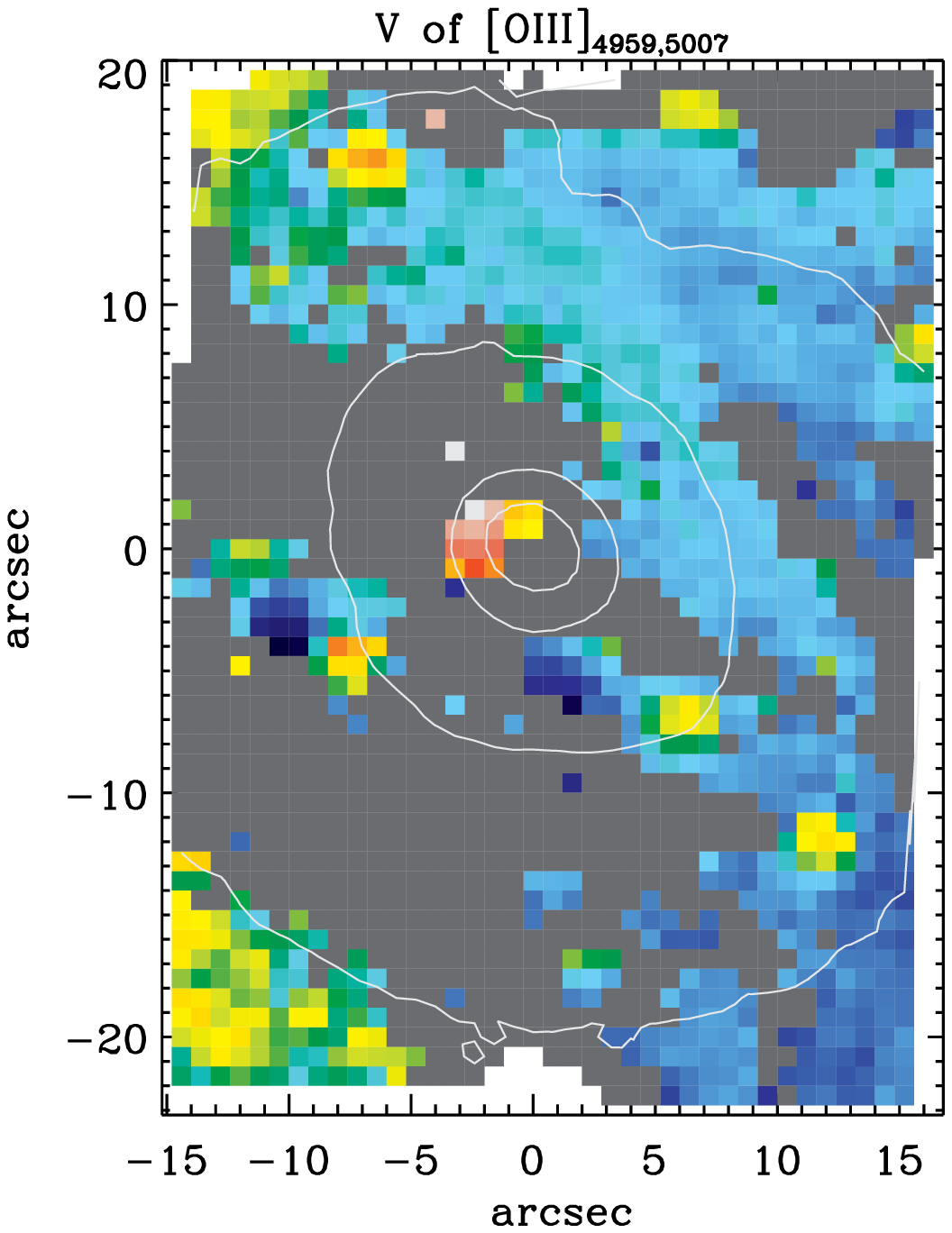}
    \end{center}
    \caption[]{Velocity of the \Oiii$\lambda\lambda4959,5007$ lines
      detected across the nuclear regions of M31, from values of
      $-300$ to 300 \kms\ relative to the systemic velocity of
      M31. Rather coherent motions are observed along the extended
      lanes of diffuse ionised-gas emission, whereas regions dominated
      by the \Oiii\ emission of known or putative PNe display
      similarly distinct velocity values. Grey areas in this and
      subsequent figures correspond to regions with no detected
      \Oiii\ emission.}
    \label{fig:VOIIImap}
  \end{figure}
}

\newcommand{\placefigFOIIImap}{
  \begin{figure}
    \begin{center}
      \includegraphics[width=\columnwidth, bb = 8 28 400 424]{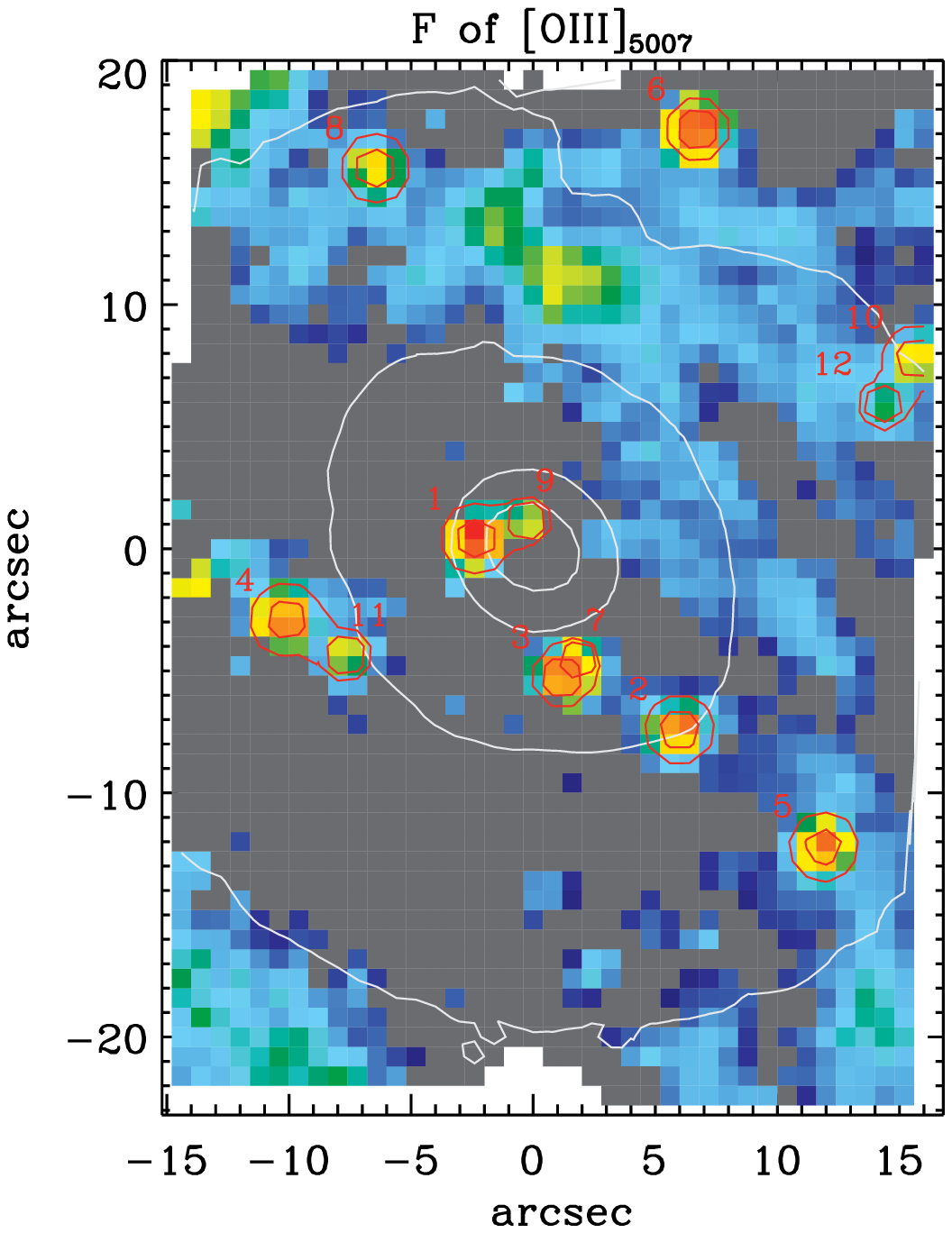}
    \end{center}
    \caption[]{Map of the flux of the \Oiii$\lambda5007$ emission from
      M31, in logarithmic scale from values of 0.5 to
      $50\times10^{-16}\rm erg\,s^{-1}cm^{-2} arcsec^{-2}$, showing
      both the diffuse ionised-gas component and each of the PNe
      considered in this paper. PNe sources are labelled and
      delineated by red contours corresponding to our best-fitting
      Gaussian model to their observed flux distribution, which is
      used to compute their total \Oiii\ flux $F_{5007}$ and
      corresponding detection limit (see text). The inner contour
      around each PNe shows the half-peak flux level of the Gaussian
      model, thus corresponding to a circle a FWHM in diameter,
      whereas the second contour shows the region containing 90\%\ of
      the flux of the model around each PN, or of both the Gaussians
      models in the case of blended PNe sources.}
    \label{fig:FOIIImap}
  \end{figure}
}

\newcommand{\placefigFluxPNefits}{
  \begin{figure*}
    \begin{center}
      \includegraphics[width=\textwidth]{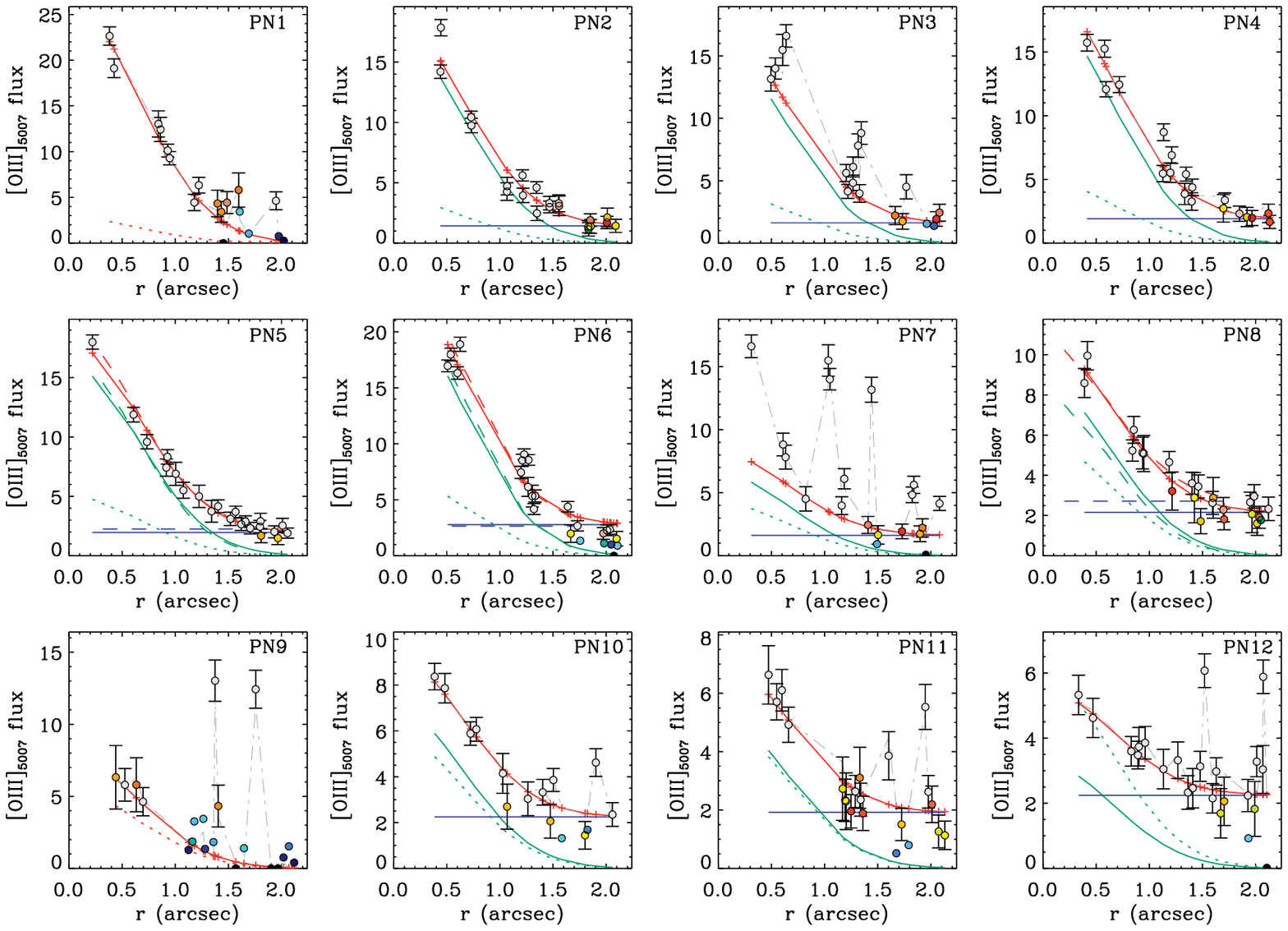}
    \end{center}
    \caption[]{Radial profiles for the \Oiii$\lambda5007$ line flux
      (in $10^{-16}\rm erg\,s^{-1}cm^{-2} arcsec^{-2}$) observed
      around the PNe shown in Fig.~\ref{fig:FOIIImap}, along with the
      corresponding best-fitting model for such a flux distribution
      (red lines, models are resampled in the
      \sauron\ $0\farcs8\times0\farcs8$ bins), which generally
      comprises a Gaussian component representing the unresolved
      emission from the PN (green lines) and a constant background
      level of diffuse ionised-gas emission (blue lines).
      Radial distances are computed from the centre of the Gaussian
      models and the data points are colour coded according to the
      value of the $A/rN$ ratio. Colours change from blue to red for
      increasing values of $A/rN$ till they saturate to white for
      $A/rN > 8$. Green corresponds to the detection threshold of
      $A/rN=4$, above which the data points are also plotted together
      with error bars for the \Oiii\ flux. The dotted coloured lines
      show the Gaussian model for the faintest PN flux that we could
      detect (see text). For PNe 12 this limit exceeds the total
      measured \Oiii\ flux, and should thus be formally regarded as
      marginal case. PNe 1 and 9, 3 and 7, 4 and 11 as well as 10 and
      12 have been modeled simultaneously, and in their corresponding
      panel the contribution of their companion PN can be appreciated
      by following the dot-dashed grey line.
      Finally, the dashed coloured lines for PNe 5, 6 and 8 show the
      models obtained through a simultaneous spectral and spatial
      fitting of the \Oiii\ emission (see text).
}
    \label{fig:FluxPNefits}
  \end{figure*}
}

\newcommand{\placefigFOIIImapHST}{
  \begin{figure}
    \begin{center}
      \includegraphics[width=\columnwidth, bb = 8 28 400 424]{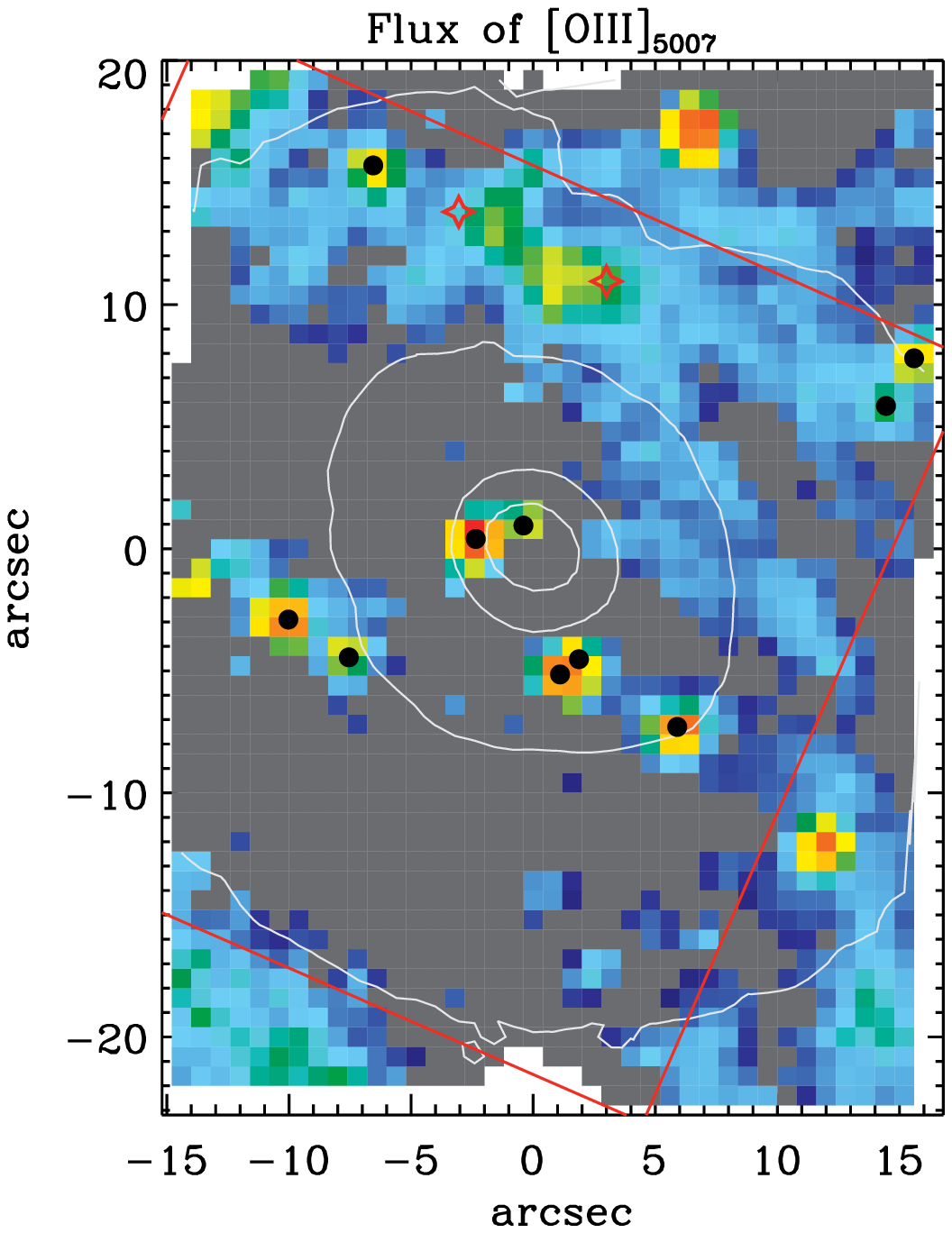}
    \end{center}
    \caption[]{Map of the flux of the \Oiii$\lambda5007$ emission from
      M31, as in Fig.~\ref{fig:FOIIImap}, but now showing the PNe
      identified in the F502N narrow-band \HST\ image (black points)
      and more specifically within the field-of-view of the WPFC2
      planetary camera (delinated by the red lines). Each of these
      sources corresponds to the PNe we isolated through our analysis
      of the \sauron\ data (Fig.~\ref{fig:FOIIImap}) except for PN~7,
      which was included in our analysis only following its
      identification in the HST images. The red crosses show the
      location of two stellar sources, most likely RGB stars.}
    \label{fig:FOIIImapHST}
  \end{figure}
}

\newcommand{\placefigFluxHSTcomparison}{
  \begin{figure}
    \begin{center}
      \includegraphics[width=\columnwidth]{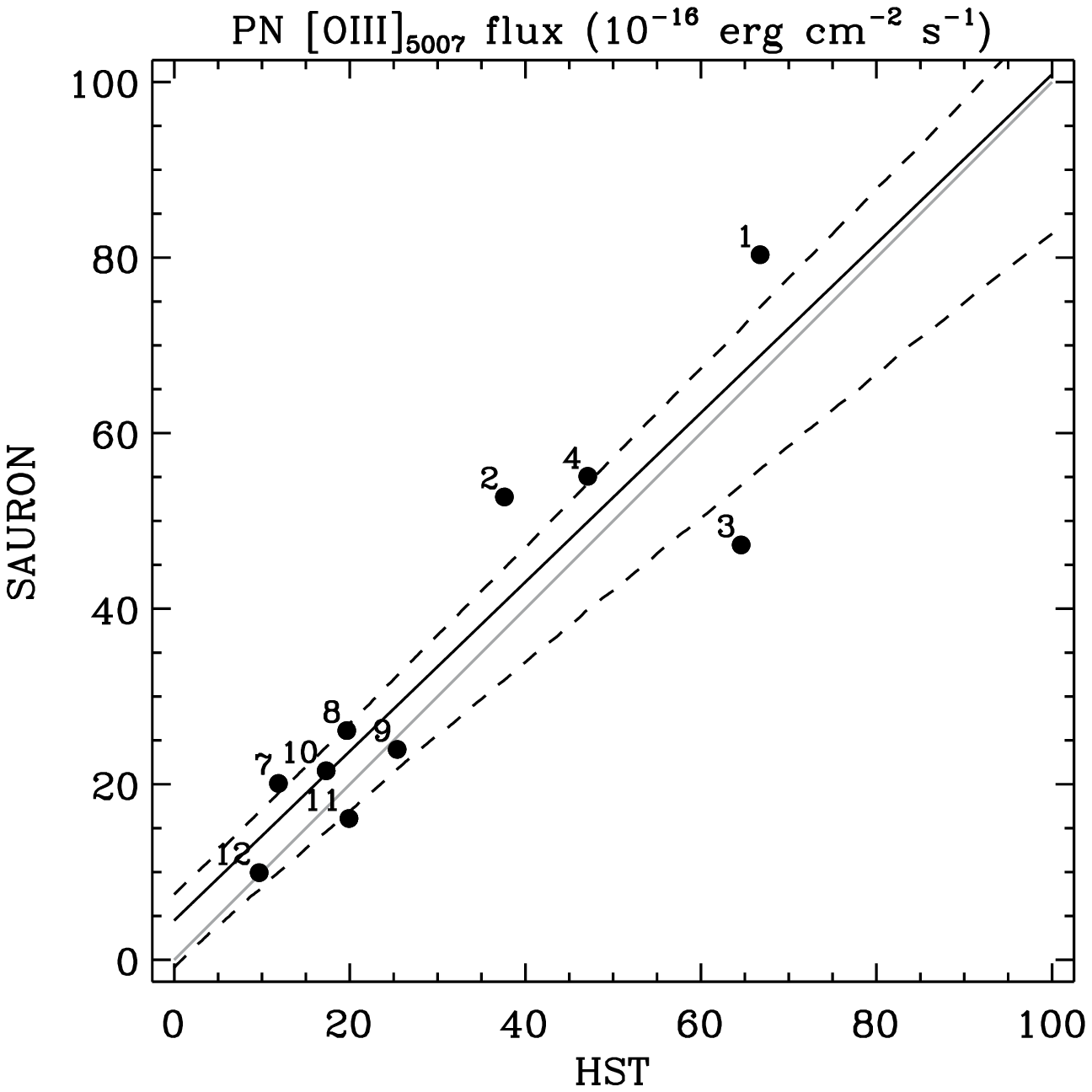}
    \end{center}
    \caption[]{Comparison between the \sauron\ and \HST\ F502N
      narrow-band measurements for the total \Oiii$\lambda5007$ flux
      of the PNe found where the field-of-view of the WFPC2-PC overlapped
      with that of the \sauron\ observations. The grey line indicates
      the one-to-one relation, whereas the solid and dashed black
      lines show the best-fitting linear regression to the data and
      associated 68\% confidence limits. The \sauron\ $F_{5007}$
      values follow rather closely their \HST\ counterparts, with no
      strong evidence of any systematic offset.}
    \label{fig:FluxHSTcomparison}
  \end{figure}
}

\newcommand{\placefigPNLF}{
  \begin{figure}
    \begin{center}
      \includegraphics[width=\columnwidth]{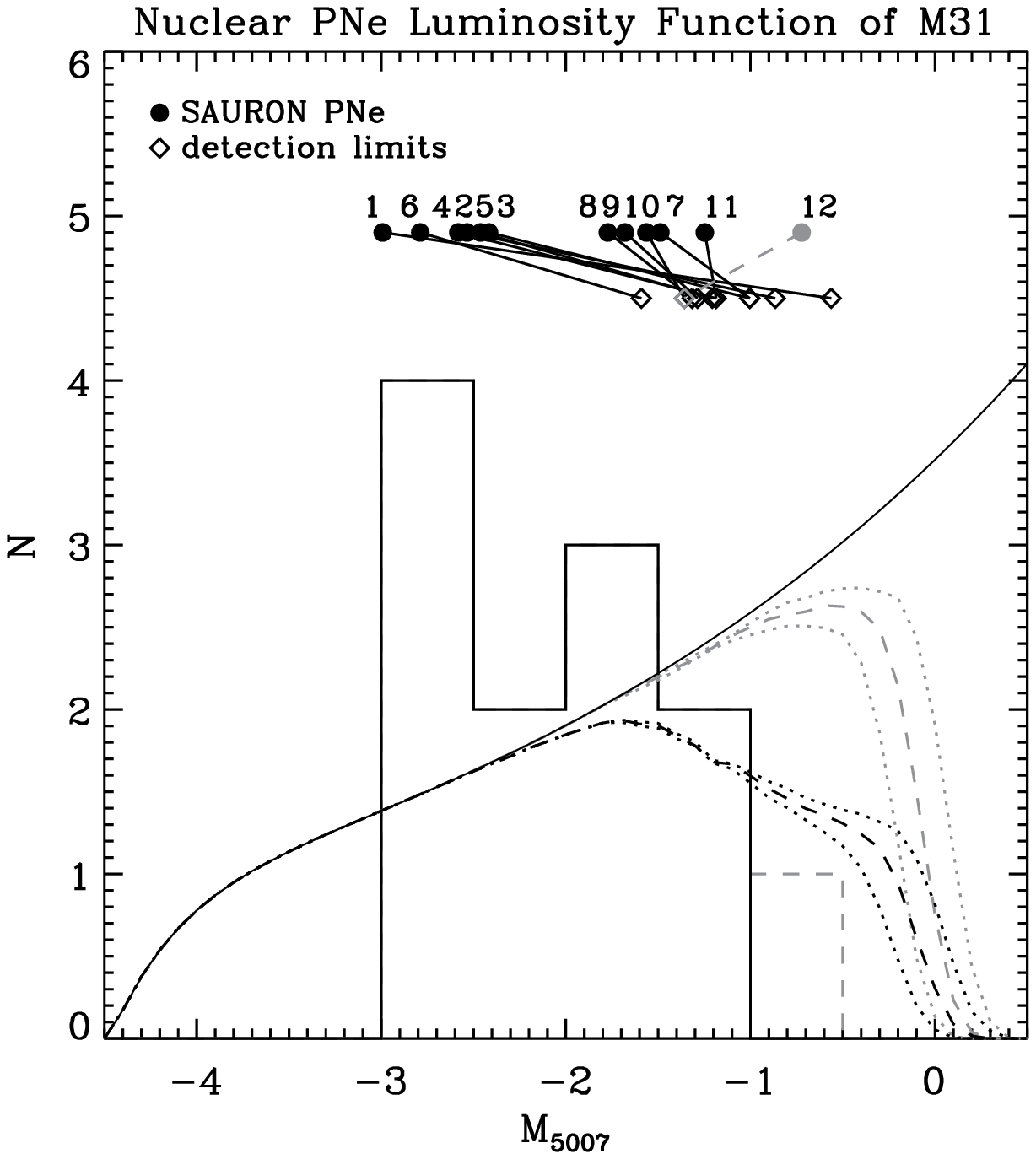}
    \end{center}
    \caption[]{Luminosity function of the PNe detected by \sauron\ in
      the central 80pc of M31 (including marginal detections, dashed
      gray histogram), along with the theoretical form of the PNe
      luminosity function \citep[solid line, from][]{Cia89}. The
      absolute magnitude of each PNe is shown at the top of the figure
      with filled circles that are plotted at an arbitrary constant
      ordinate and that are connected to open diamonds showing the
      detection limit of each source (with the marginal source PN~12
      shown in grey).
      The dashed line shows the PNe luminosity function multiplied by
      the median values of the completeness function across the entire
      \sauron\ field, whereas the dotted lines indicate the range by
      which the expected number of detected PNe would vary depending
      on their exact position within the $0\farcs8\times0\farcs8$
      \sauron\ bins.
      Such a completeness function accounts also for the presence of
      diffuse ionised-gas in the central regions of M31, whereas the
      dashed and dotted grey lines show the impact of ignoring such a
      component.
      Even though a Kolgomorov-Smirnov test indicates only a 46\%
      probability that the observed PNe luminosity distribution was
      drawn from the completeness-corrected theoretical form of the
      PNe luminosity function, here we still normalised the latter so
      that it would lead to an expected number of detected PNe that
      matches the one we observe (that is 12, including marginal
      detections).
}
    \label{fig:PNLF}
  \end{figure}
}

\newcommand{\placefigCompleteness}{
  \begin{figure*}
    \begin{center}
      \includegraphics[width=\textwidth]{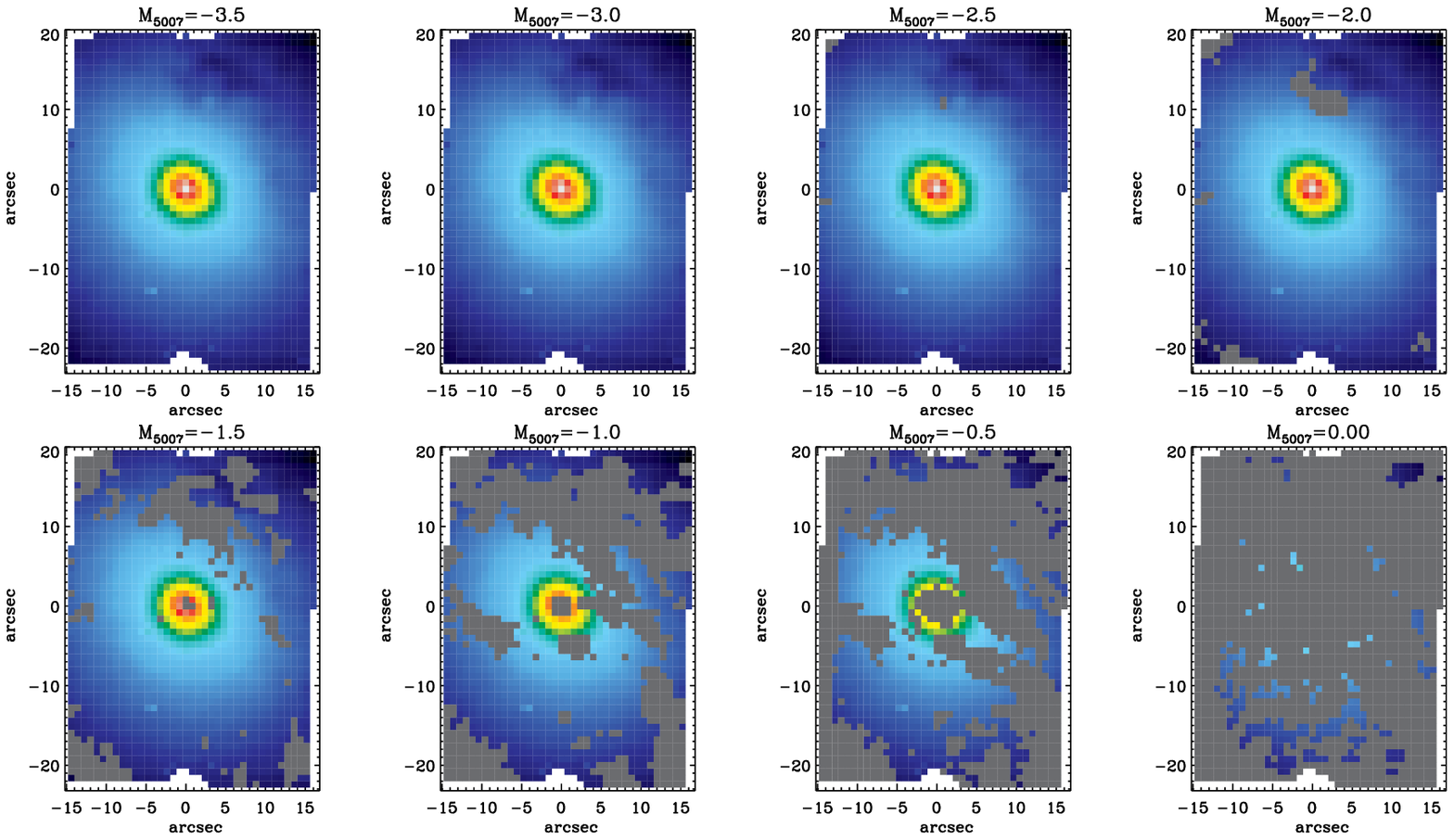}
    \end{center}
    \caption[]{Reconstructed optical images of M31, in logarithmic
      flux scale, showing with grey bins the regions where, from left
      to right, PNe of decreasing luminosity would escape
      detection. To check whether in a given \sauron\ bin a PN of
      absolute magnitude $M_{5007}$ would be detected, we generated at
      that position a Gaussian model for the \Oiii\ flux of a PN of
      that brightness at the distance of M31, deriving also the
      corresponding spectral density values for the amplitude of the
      \Oiii\ line. Using the values for the level of the noise $rN$ in
      the residuals of our spectral fits we then computed the values
      of the $A/rN$ ratio around the PN position and, using the
      criterion introduced in \S~\ref{subsec:AnalysisPNeDetection},
      simply checked whether $A/rN > 4$ within a FWHM from the centre
      of the Gaussian model, which in these particular simulations
      correspond to the centre of the \sauron\ bins. Furthermore, in
      the presence of diffuse ionised-gas emission we also required
      that the peak \Oiii\ flux for the simulated PNe exceeds by more
      than three times the diffuse nebular flux, where the latter was
      taken to be the background level estimated during our spatial
      fits (see Fig~\ref{fig:FluxPNefits}) in regions where a PN was
      actually found in our data.
      Since the number of PNe scales with the number of stars, at any
      $M_{5007}$ value the ratio of the stellar flux observed where
      PNe of that absolute magnitude can be detected over the total
      stellar flux encompassed by our field-of-view corresponds the
      probability that a PN of that $M_{5007}$ value could be detected
      during our observations, thus serving to construct our
      completness function. 
      Uncertainties on such a function can be estimated by randomly
      placing the PNe Gaussian models within each
      \sauron\ $0\farcs8\times0\farcs8$ bin, rather than exactly at
      their centres.}
    \label{fig:Completeness}
  \end{figure*}
}

\newcommand{\placefigStellarPop}{
  \begin{figure}
    \begin{center}
      \includegraphics[width=\columnwidth]{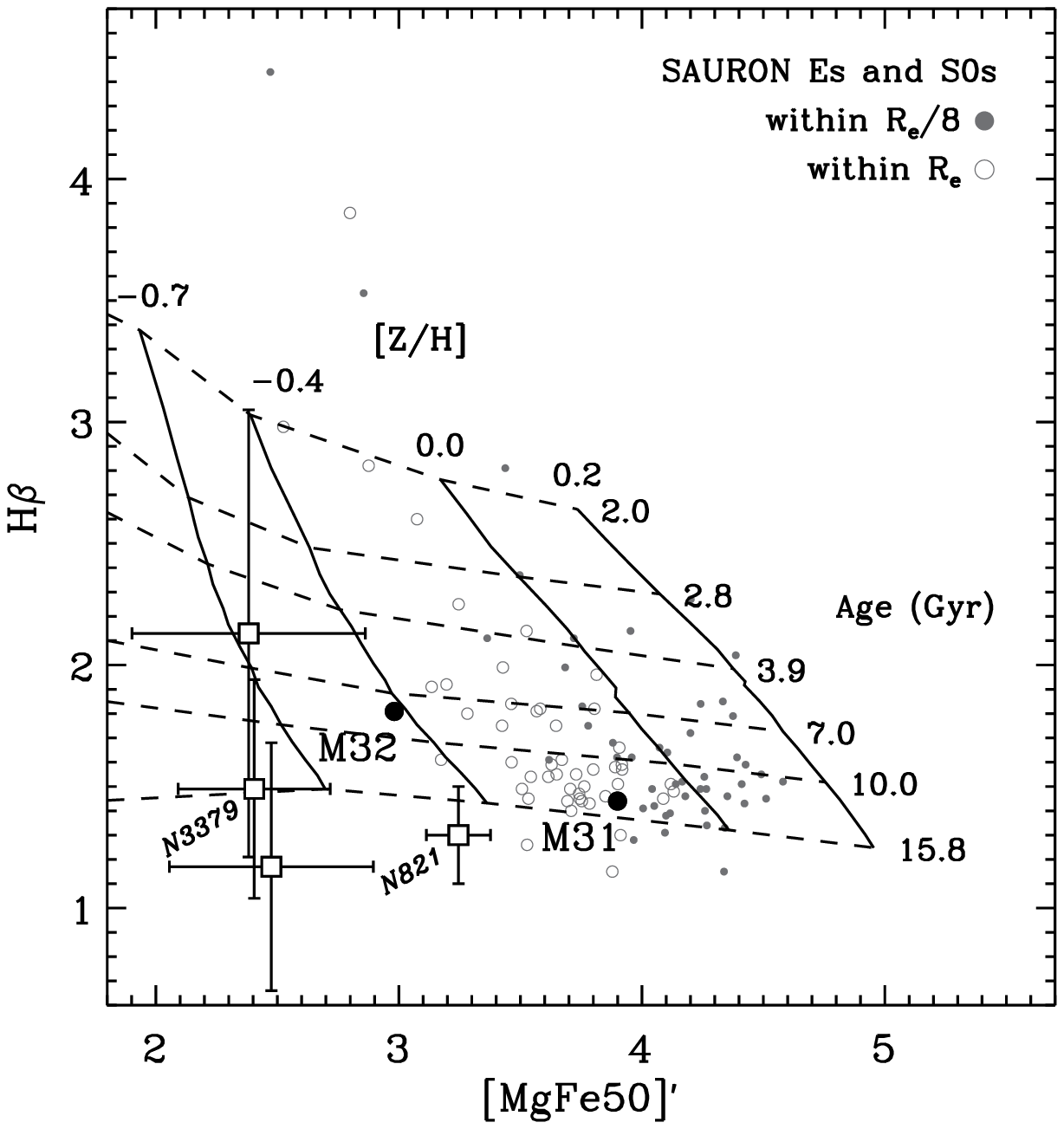}
    \end{center}
    \caption[]{The \Hb\ vs [MgFe50'] diagnostic diagram of
      \citet{Kun10} where the stellar age and metallicity of the
      central 80 pc of M31 can be compared with i) that of the optical
      regions of M32 (i.e., within its effective radius $R_e$), ii)
      that of the 48 early-type galaxies of the \sauron\ sample
      (within both one $R_e$ and $R_e/8$ as shown by the open and
      filled grey circles, respectively; from \citealp{Kun06}) and
      iii) with what found using the \sauron\ spectrograph in the
      halos of NGC~3379 and NGC~821 (open squares, from the
      measurements of \citealp{Wei09} centred between 3 and 4
      $R_e$ for NGC~3379 and at around one $R_e$ for NGC~821).
      The age and metallicity grids are from the models of
      \citet{Sch07}, as adapted for the study of \citet{Kun10}.
      The central stellar population of the \sauron\ early-type
      galaxies tend to be more metal-rich than what is observed in the
      nuclear regions of M31, which on the other hand display a
      stellar metallicity value that generally exceeds what found
      within the whole optical regions of early-type galaxies.
      In fact, that would appear to be certainly the case of low-mass
      objects such as M32, which sports a value for the stellar
      metallicity within its optical regions that is more akin,
      although not as low, to what is observed in the halos of early-type
      galaxies.}
    \label{fig:M31pop}
  \end{figure}
}

\newcommand{\placetabone}{
\begin{table*}
\caption{Basic Properties of the PNe in the Optical Regions of M31}
\label{tab:PNeProp}
\begin{center}
\begin{tabular}{rrrrrrrrrrrr}
\hline
ID  & x-off  & y-off  & $F_{5007}$ & $\delta F_{\rm 5007}$ & $F_{\rm 5007, lim\/}$ & $F_{\rm 5007, spec\/}$ & $F_{5007, HST\/}$ & $V_{\rm PN}$ & $V_{\star}$ & $\sigma_{\star}$ & \Oiii/\Hb \\ 
(1) & (2)    & (3)    & (4)        & (5)                   & (6)                   & (7)                    & (8)               & (9)          & (10)        & (11)             & (12)      \\
\hline
 1  &  -2.35 &   0.42 & 80.3       &  3.1                  &  8.6                  &  ---                   & 66.7              &  262.3       &  57.3       & 120.4            & $>15.1$   \\     
 2  &   6.00 &  -7.39 & 52.7       &  2.8                  & 11.3                  &  ---                   & 37.6              &   44.4       & -24.3       & 153.1            &    8.0    \\   
 3  &   1.17 &  -5.28 & 47.3       & 10.9                  & 12.9                  &  ---                   & 64.6              & -183.4       &   0.5       & 140.3            & $>12.7$   \\     
 4  & -10.10 &  -2.91 & 55.1       &  2.4                  & 15.2                  &  ---                   & 47.1              & -214.1       &  21.8       & 144.5            &   17.2    \\    
 5  &  11.91 & -12.20 & 49.3       &  1.5                  & 15.5                  & 46.4                   &  ---              &   61.2       & -31.2       & 155.1            &    7.4    \\ 
 6  &   6.74 &  17.18 & 66.6       &  4.7                  & 22.1                  & 66.4                   &  ---              &   13.1       &  10.6       & 176.0            &   11.5    \\ 
 7  &   1.81 &  -4.57 & 20.1       & 10.1                  & 12.9                  &  ---                   & 11.9              & -127.1       &  -7.8       & 144.2            & $>10.2$   \\  
 8  &  -6.45 &  15.59 & 26.1       &  1.9                  & 17.2                  & 22.1                   & 19.7              &  110.5       &  18.3       & 139.4            &    7.4    \\ 
 9  &  -0.27 &   1.15 & 24.0       &  5.3                  & 16.7                  &  ---                   & 25.4              &   85.1       &  82.6       & 167.9            &  $>2.8$   \\  
10  &  15.65 &   7.83 & 21.5       &  1.4                  & 17.8                  &  ---                   & 17.3              &   25.1       & -16.1       & 160.4            &    5.2    \\
11  &  -7.68 &  -4.35 & 16.1       &  2.8                  & 15.2                  &  ---                   & 19.9              &  108.1       &   6.7       & 138.7            &  $>7.9$   \\ 
12  &  14.36 &   5.93 &  9.9       &  1.4                  & 17.8                  &  ---                   &  9.7              &  -50.6       &  -9.9       & 157.3            &    5.0    \\
\hline
\end{tabular}
\end{center}
\begin{minipage}{\textwidth}
{\it Notes:\/}
(1)~PN ID.  
(2)--(3)~R.A. and Declination offset position, in arcseconds, from the
  centre of M31.
(4)--(6)~Total \Oiii$\lambda5007$ flux $F_{5007}$ in $10^{-16}\rm
  erg\,s^{-1}cm^{-2}$, with associated formal error and detection
  limit.
(7)--(8)~Same as Col. (4) but now for $F_{5007}$ values measured
through a simultaneous spectral and spatial fit of the \sauron\ data,
and from WFPC2 narrow-band images in the F502N filter.
(9)~Velocity of the PN, in \kms\ and relative to the systemic velocity
of M31, measured in spectra extracted within a FWHM-wide aperture
around the center of the best-fitting Gaussian models shown in
Figs.~\ref{fig:FOIIImap} and \ref{fig:FluxPNefits}.
(10)--(11)~Velocity and velocity dispersion of the stars along the
line-of-sight pointing to the direction of the PN, measured from the
same aperture spectra.
(12)~\OiiioHb\ ratio, again in the same spectra. An upper-limit is
listed when \Hb\ is not detected, using an \Hb\ flux that would
correspond to lines with the same profile as the \Oiii\ lines and with
an amplitude equal to 3 times the level of noise in the fit residuals
(i.e., the detection threshold for \Hb, following \citealp{Sar06}).
\end{minipage}
\end{table*}
}

\title[PNe in the nuclear regions of M31]{The Planetary Nebulae
  Population in the Nuclear Regions of M31: the SAURON view}

\author[Pastorello et al.]{Nicola Pastorello$^{1}$\thanks{E-mail:
    npastorello@swin.edu.au}, Marc Sarzi$^{2}$, Michele
  Cappellari$^{3}$, Eric Emsellem$^{4,5}$, \newauthor Gary A.\ Mamon$^{6}$, 
  Roland Bacon$^{5}$, Roger L.\ Davies$^{3}$, P.~Tim de
  Zeeuw$^{4,7}$\\
$^{1}$Centre for Astrophysics and Supercomputing, Swinburne University, 
Hawthorn, VIC 3122, Australia \\
$^{2}$Centre for Astrophysics Research, University of Hertfordshire,
College Lane, Hatfield, Herts, AL10 9AB, UK\\
$^{3}$Sub-Dept of Astrophysics, Dept of Physics, University of Oxford,
Denys Wilkinson Building, Keble Road, Oxford, OX1 3RH, UK \\
$^{4}$European Southern Observatory, Karl-Schwarzschild-Str~2, 85748
Garching, Germany\\
$^{5}$Centre de Recherche Astronomique de Lyon, 9~Avenue Charles
Andr\'e, 69230 Saint Genis Laval, France\\
$^{6}$Institut d'Astrophysique de Paris (UMR 7095: CNRS \& UPMC), 98
bis Bd Arago, 75014 Paris, France\\
$^{7}$Sterrewacht Leiden, Universiteit Leiden, Postbus 9513, 2300 RA
Leiden, The Netherlands }

\begin{document}
\pagerange{\pageref{firstpage}--\pageref{lastpage}} \pubyear{2010}

\maketitle
\label{firstpage}

%
\begin{abstract}

The study of extragalactic Planetary Nebulae (PNe) in the optical
regions of galaxies, where the properties of their stellar population
can be best characterised, is a promising ground to better understand
the late evolution of stars across different galactic environments.
Following a first study of the central regions of M32 that illustrated
the power of integral-field spectroscopy (IFS) in detecting and
measuring the \Oiii$\lambda5007$ emission of PNe against a strong
stellar background, we turn to the very nuclear PN population of M31,
within $\sim$80 pc of its centre.
We show that PNe can also be found in the presence of emission from
diffuse gas, as commonly observed in early-type galaxies and in the
bulge of spirals, and further illustrate the excellent sensitivity of
IFS in detecting extragalactic PNe through a comparison with
narrow-band images obtained with the Hubble Space Telescope.
Contrary to the case of the central regions of M32, the nuclear PNe
population of M31 is only marginally consistent with the generally
adopted form of the PNe luminosity function (PNLF). In particular,
this is due to a lack of PNe with absolute magnitude $M_{5007}$
brighter than $-3$, which would only result from a rather unfortunate
draw from such a model PNLF.
The nuclear stellar population of M31 is quite different from that of
the central regions of M32, which is characterised in particular by a
larger metallicity and a remarkable UV-upturn. We suggest that the
observed lack of bright PNe in the nuclear regions of M31 is due to a
horizontal-branch population that is more tilted toward less massive
and hotter He-burning stars, so that its progeny consists mostly of
UV-bright stars that fail to climb back up the asymptotic giant branch
(AGB) and only of few, if any, bright PNe powered by central post-AGB
stars.
These results are also consistent with recent reports on a dearth of
bright post-AGB stars towards the nucleus of M31, and lend further
support to the idea that the metallicity of a stellar population has
an impact on the way the horizontal branch is populated and to the
loose anticorrelation between the strength of the UV-upturn and the
specific number of PNe that is observed in early-type galaxies.
Finally, our investigation also serves to stress the importance of
considering the same spatial scales when comparing the PNe population
of galaxies with the properties of their stellar populations.

\end{abstract}

%

%
\section{Introduction}
\label{sec:intro}

Planetary Nebulae (PNe) in external galaxies are mostly regarded
either as tracers of the gravitational potential
\citep[e.g.,][]{Rom03,Dou07} or as indicators for the distance of
their galactic hosts \citep[e.g.,][]{Cia89,Jac90,Jac92}, with the
latter advantage owing to the nearly universal -- though not fully
understood -- shape of the PNe luminosity function (PNLF, generally in
the \Oiii$\lambda5007$ emission).
Yet extra-galactic PNe can also be used as probes of their parent
stellar population \citep[e.g.,][]{Ric99,Jac99,Dop97} and
understanding in particular the origin of the PNLF is a puzzle that,
once solved, promises to reveal new clues on the late stages of
stellar evolution and on the formation of PNe themselves \citep[see,
  e.g.,][]{Cia06}.

PNe originate from horizontal-branch (HB) stars that climb back up the
asymptotic giant branch (AGB) at the end of their helium-burning
phase, after which these stars leave the AGB and quickly cross the
Hertzprung-Russell diagram on their way towards the cooling track of
white dwarves (WD).
For a population with a given age and metallicity, HB stars have
nearly the same helium core mass ($\sim 0.5\,M_{\odot}$) but a range
of hydrogen shell mass ($\sim 0.001 - 0.3 \,M_{\odot}$), with the
reddest stars having also the largest H-shells and originating from
the most massive main-sequence progenitors.
Only HB stars with a considerable H-shell ascend toward the AGB and
eventually lead to the formation of a PN (being known at this point as
post-AGB or early post-AGB stars depending on their UV brightness),
whereas the bluest HB stars with little envelope mass head straight
toward the WD cooling curve by evolving first to higher luminosities
and effective temperatures (the so-called AGB-manqu\'e phase, see
\citealp{Gre99}).

According to this simple picture, galaxies with on-going star
formation should show brighter PNe than quiescent systems where
massive stars have long disappeared (e.g., \citealp[][]{Mar04} show
that the most luminous PNe arise from 1 Gyr old 2.5 $M_\odot$ stars),
but in fact the PNLF of young and old galaxies are relatively similar.
In particular, all extra-galactic PNe surveys indicate a common and
bright cut-off for the PNLF, which led \citet{Cia05} to suggest a
binary evolution for the progenitors of the brightest PNe that would
be common to different kind of galaxies.
Alternatively, a similarly bright PNLF cutoff in elliptical galaxies
might be explained by minor mergers mixing in 1 Gyr old stars with the
much older stellar population of these galaxies \citep{Mam05}.
If galaxies seem to invariably host very bright PNe, their specific
content of PNe - that is the number of PNe normalised by a galaxy
bolometric luminosity - appears to vary with the metallicity of the
stellar population. More specifically, \citet{Buz06} found that more
metal-rich galaxies show comparably fewer PNe, which also corresponds
to larger far-UV fluxes.
Interestingly, this may indicate that, at a given mean stellar age, a
larger metallicity would bias the HB population towards fewer stars
with a massive H-shell capable of leading to the formation of PNe.  On
the other side, a larger fraction of blue HB stars will instead
contribute to the overall far-UV flux of their host galaxy
\citep[i.e. the so-called UV-upturn,][]{Bur88} as they follow their
AGB-manqu\'e evolution.
This trend is not completely unexpected given that the mass of the
H-shell around HB stars depends on the amount of mass they have lost
on the red-giant branch and that in turn the mass-loss rate efficiency
$\eta$ should increase with stellar metallicity \citep{Gre90}.
Yet, it must be borne in mind that $\eta$ could also follow an
increase in the abundance of Helium that may also come with a larger
stellar metallicity.

Within this context, we note that whereas our knowledge of both the
shape and normalisation of the PNLF comes chiefly from the peripheral
PN populations of galaxies, measurements of both the stellar
metallicity and the UV spectral shape of galaxies pertain to their
optical regions.
Indeed, with ground-based narrow-band imaging or slitless spectroscopy
- the most common techniques employed to find extragalactic PNe - is
is difficult to detect PNe against a strong stellar background,
whereas measuring the strength of stellar absorption lines or imaging
the far-UV flux of galactic halos is prohibitively expensive in terms
of telescope time.
Such a dramatic spatial inconsistency needs to be resolved to
understand the link between PNe and the properties of their parent
stellar populations, in particular if we consider that such a
connection may already not be entirely within our grasp, as suggested
by the Hubble Space Telescope (\HST) observations of M32 obtained by
\citet{Bro08} who uncovered a dearth of UV-bright stars compared to
what expected from stellar evolutionary models.

In our investigation of the compact elliptical M32 \citep[][hereafter
  Paper~I]{Sar11} we have demonstrated how integral-field spectroscopy
can overcome the previous limitations and detect PNe in the central
regions of galaxies.
In fact, with \sauron\ data taken with just two 10-minute pointings
we could double the number of known PNe within the effective radius of
M32 and detect PNe five times fainter than previously found in
narrow-band images that collected nearly the same number of photons.
Here we will turn our attention to the nuclear regions of M31, using
\sauron\ observations of similar depth and spatial coverage, and
motivated by the known differences between the stellar metallicity and
UV colours of M31 and M32 \citep[e.g.,][]{Bur88}.

This paper is structured as follows. In \S\ref{sec:Data}, we briefly
review the acquisition and the reduction of the \sauron\ data for M31.
In \S\ref{sec:Analysis}, we detail our method for identifying PNe and
measuring their \Oiii$\lambda5007$ flux, accounting also for the
presence of diffuse ionised-gas emission in M31 and comparing our PN
detections and flux measurements with the ones obtained from
\HST\ narrow-band imaging.
In \S\ref{sec:Results}, we assess whether our data are consistent with
the generally adopted shape for the PNLF, finding evidence for a
dearth of bright PNe in the nuclear regions of M31, which we then
discuss in \S\ref{sec:Discussion} in the context of the stellar
population properties of this and other galaxies.
Finally, in \S\ref{sec:Conclusions} we draw our conclusions.

Throughout this paper we assume a distance of 791 kpc for M31, taken
to be equal to that to M32 based on the surface-brightness fluctuation
measurements of \citet{Ton01}.

\section{Observations and Data Reductions}
\label{sec:Data}

M31 (NGC~224) was one of the special objects that were observed over
the course of the \sauron\ representative survey \citep{deZ02}. Its
central regions were observed with two 1800s slightly offset pointings
and using the low-resolution mode of \sauron, which gives a
$33\farcs0\times44\farcs0$ field-of-view fully sampled by
$0\farcs94\times0\farcs94$ square lenses \citep[for more details on
  the instrument see][]{Bac01a}.
At the assumed distance of 791 kpc such an field corresponds to a
circular area 76 pc in radius.
The data from each pointing were reduced similarly to the data
obtained for the objects of the main \sauron\ sample
\citep[see][]{Ems04,Fal06}, and the resulting datacubes were merged
and resampled in $0\farcs8\times0\farcs8$ spatial elements each
corresponding to spectra covering the wavelength range between
4830\AA\ and 5330\AA\ with a final spectral resolution of
4.2\AA\ (FWHM, for a $\sigma_{\rm res}=108\kms$).
The only differences with the data used in the papers of the
\sauron\ project are that here we did not perform any Voronoi spatial
binning \citep{Cap03}, to avoid swamping the signal of the weaker PNe
against an increased stellar background, and that we re-adjusted the
absolute flux calibration of the \sauron\ cube using data obtained
with \HST.

In particular, we have used archival Wide-Field Planetary Camera
(WFPC2) narrow-band images obtained in the F502N passband, since this
filter falls entirely within the wavelength range of our \sauron\ data
and because the typical old stellar spectrum of early-type galaxies
and bulges can be considered pretty much as a flat spectral energy
density source across the F502N filter.
In fact, since the {\tt PHOTFLAM\/} keyword in the \HST\ images is
defined as the flux density of a flat spectral source (in $\rm
erg\,s^{-1}cm^{-2}\AA^{-1}$) that would produce a flux corresponding
to one count per second, this situation allows to use the {\tt
  PHOTFLAM\/} keyword to construct a flux-density profile that can be
readily matched (accounting for the different pixel size) to the
radial trend for the median flux density of our \sauron\ spectra
(where such a median value is not affected by the presence of
\Oiii\ emission).

\placefigNarrowBand
\placefigAoNmap
\placefigVOIIImap

%
\section{Data Analysis}
\label{sec:Analysis}

\subsection{Emission-line Measurements}
\label{subsec:AnalysisEmission}

In order to identify the PNe in the nuclear regions of M31 and measure
their flux in the \Oiii$\lambda5007$ line, we first need to separate
as accurately as possible the stellar and nebular contribution to each
of the \sauron\ spectra, and then to further disantangle the emission
due to PNe from the one arising from diffuse gas.

Extended ionised-gas emission is common in early-type galaxies and in
the bulges of disc galaxies \citep{Sar06,Fal06}.
A first \Oiii\ narrow-band image made from the SAURON data
(Fig.~\ref{fig:NarrowBandmap}, see caption for details) indeed reveals
prominent lanes of diffuse emission, as well as a number of compact
sources of \Oiii\ emission, some of which were already recognised as
PNe by \citet{Cia89} and \citet{Mer06}.

In principle, the PNe and diffuse ionised-gas emission have
sufficiently distinct characteristics to produce complicated line
profiles where both components occur, such that properly extracting
the flux of a PN surrounded by diffuse emission may require to
separate the relative contribution of these components directly in the
spectra (for instance through a double-Gaussian fit).
In fact, across a given region where a putative PN would spread its
nebular flux according to the telescope and seeing point-spread
function (PSF), the PNe emission lines should be spectrally unresolved
in our spectra (PNe have an intrinsic line profile with a velocity
dispersion of only of few tens of \kms, much smaller than our
instrumental resolution), falling always at the position corresponding
to the velocity of the PN and with a strength - or peak spectral
density amplitude - that varies between different spatial bins
according to the shape of the PSF (e.g. a bi-dimensional
Gaussian). The putative PN spectrum should also be always
characterised by the same \Oiii/\Hb\ line ratio.
On the other hand, the diffuse ionised-gas emission could, in
principle, come with different kinematics, strength and line-ratios
across the entire region where the PN emission is found.

In practice, however, we do not expect to observe dramatic variations
in the properties of the diffuse gas where the light of a given PN
falls, either because such fluctuations are not likely to occur across
the physical scale that correspond to a seeing disc (that is,
$\sim\,$14 pc for a $\sigma_{\rm PSF} = 0\farcs61$ - see \S3.2 for how
this is derived) or since they will be considerably levelled by the
PSF itself anyway (for instance, see Fig.~\ref{fig:VOIIImap} below for
an illustration of how smooth the motion of the diffuse component is).
Furthermore, the velocity of the PNe in the nuclear regions of M31 is
unlikely to differ from that of the diffuse ionised-gas by more than a
few 100 \kms\, given the modest value for the stellar velocity
dispersion \citep[$\sim$160 \kms;][]{Sag10} and that both gas and
stars rotate relatively slowly (at $\sim$ 50 and 100 \kms,
respectively).
This means that the observed line profiles should not be exceedingly
complicated in the presence of both PN and diffuse gas components
(given the limited spectra resolution of our data), remaining in
particular in the regime where a single Gaussian profile estimates
sufficiently well the {\it total\/} nebular flux in our
spectra\footnote{Simple tests show that a single Gaussian fit to
  double-Gaussian profiles allow to estimate to total flux with less
  than 10\% uncertainty as long as the two Gaussian components are
  separated by less than 3 times their common dispersion. This is
  irrespective of the relative strength of the two Gaussians but
  assumes similar values for their dispersion, which is justified for
  our case study where both the PNe and the diffuse gas emission are
  characterised by intrinsic profiles that are hardly resolved.} and
thus allowing us to subsequently discern the presence of PNe using
only a spatial analysis of the \Oiii\ flux distribution.

For these reasons we proceeded to extract the ionised-gas flux values
in our spectra by adopting the fitting method of \citet{Sar06},
whereby a set of stellar templates and Gaussian emission lines are
fitted simultaneously to the spectra\footnote{
In practice this is achieved by using the IDL code GANDALF (available
at {\tt http://star-www.herts.ac.uk/$_{\widetilde{~}}$sarzi\/}) and
the stellar kinematics extracted with the pixel-fitting IDL code pPXF
\citep[][{\tt
    http://www-astro.ox.ac.uk/$_{\widetilde{~}}$mxc/idl\/}]{Cap04}
}, while adopting the approach of
\citet{Sar10} to further improve the match to the stellar continuum
and ensure that the ionised-gas emission is extracted from the
subtraction of a physically motivated stellar model.

\placefigFOIIImap
\placefigFluxPNefits
\placetabone

As a first result of this procedure, Fig.~\ref{fig:AoNmap} shows the
map of the ratio between the amplitude $A$ of the best-fitting
Gaussian to the \Oiii$\lambda5007$ line and the noise level $rN$ in
the residuals of the overall fit. The value of the $A/rN$ ratio
relates to the accuracy with which the flux, position and width of a
single emission line can be estimated, with tests indicating a formal
detection for $A/rN > 4$ \citep{Sar06}.
Dark blue colours in Fig.~\ref{fig:AoNmap} show regions with no
detected \Oiii\ emission (hereafter shown in grey) and where
high-quality SAURON aperture spectra were extracted and subsequently
converted into stellar templates in order to improve our spectral fit
\citep{Sar10}.

Compared to Fig.~\ref{fig:NarrowBandmap}, the known PNe in the nuclear
regions of M31 show up with a greater contrast as unresolved sources
in the $A/rN$ of Fig.~\ref{fig:AoNmap}, which further reveals the
extent of the diffuse ionised-gas emission and the presence of the
central PN already detected with integral-field data by \citet{Bac01b}
A few additional unresolved sources could be present in
Fig.~\ref{fig:AoNmap}, for instance around 8\arcsec\ east of the
nucleus and next to PN C9 of \citet{Cia89} or next the western edge of
the \sauron\ field-of-view and around 8\arcsec\ north of the nucleus.

The kinematics of the \Oiii\ lines provides more elements to judge
whether a given patch of \Oiii\ is likely to originate from a PN,
since PNe are expected to generally move with a different velocity
than the gas clouds responsible for the diffuse \Oiii\ emission that
may be found along the same line of sight.
Fig~\ref{fig:VOIIImap} shows indeed that where the known PNe dominate
the nebular emission, the \Oiii\ kinematics appears quite coherent and
at the same time significantly different from that of the surrounding
diffuse component when this is present. This seems to be the case also
for the two aforementioned putative sources, with the possible
addition of few more regions.

\subsection{PNe Detection and Flux Measurements}
\label{subsec:AnalysisPNeDetection}

In order to measure the total \Oiii\ emission of the known PNe in the
nuclear regions of M31 and assess more precisely the presence of
additional sources, we proceeded to fit the \Oiii\ flux distribution
around any putative PNe with a bi-dimensional Gaussian profile meant
to represent the PSF, like already detailed in Paper~I (including the
possibility to match blended sources). 
Furthermore, we now also allow for a constant background level to
account for the diffuse ionised-gas emission.
During each of these fits, we adjusted the position and amplitude of
the Gaussian models as well as the level of the background, whereas
the full-width at half maximum (FWHM) of the PSF was optimized over
the whole set of sources that were eventually deemed detected. This
turned out to have a value of 1\farcs44, for a $\sigma_{\rm
  PSF\/}=0\farcs61$.
Considering a constant level for the background diffuse emission
certainly represents a simple approximation and yet, despite it being
meant initially only as a starting point for our modelling, this
approach eventually turned out to work as well as more sophisticated
models (see below).

In our previous analysis of M32 (Paper~I), we deemed a given source
that was successfully matched by our model a detected PNe if all
\Oiii\ lines are detected (i.e., $A/rN > 4$ in the present case)
within at least the FWHM of the bi-dimensional Gaussian model.
The presence of diffuse ionised-gas in M31 has led us to require in
addition that the peak amplitude of the Gaussian model exceeds by at
least a factor 3 the background level of the diffuse emission, when
this is needed.
Furthermore, we should make sure that where the nebular emission is
dominated by our putative PNe (i.e., within a FWHM), the values for
the \OiiioHb\ ratio significantly exceed the typical values observed
in the filament structures in the nuclear regions of M31, which range
between 1 and 3, similarly to the case of the diffuse emission of
early-type galaxies \citep{Sar06}.

All the PNe that we eventually detected according to these criteria
are located in the map for the \Oiii$\lambda5007$ flux shown in
Fig.~\ref{fig:FOIIImap}. One of these sources (PN 7) was only
identified following our analysis of \HST\ images (see next section)
and another (PN 12) represents only a marginal detection as the peak
amplitude of its best-fitting bi-dimensional Gaussian is only twice
the level of the background emission.
Fig.~\ref{fig:FOIIImap} also nicely illustrates the relatively smooth
character of the diffuse emission observed in the filamentary
structures in central regions of M31, which supports our simple
treatment of this component in our PNe fit as a constant background.

The quality of our fit to the \Oiii\ flux distribution of our detected
PNe sources and their immediate surroundings can also be appreciated
in Fig.~\ref{fig:FluxPNefits}, where the radial profile for the
\Oiii\ flux of each PN is plotted together with its best-fitting
Gaussian model.
Fig.~\ref{fig:FluxPNefits} also shows the Gaussian profile that
corresponds to the limiting total PN flux that we could have measured,
which is obtained by scaling down the best fitting Gaussian model
until either its peak amplitude becomes less than three times the
level of diffuse-gas emission (when present) or until it would induce
the faintest \Oiii\ line within a FWHM to become undetected.

For all detected PNe, Tab.~\ref{tab:PNeProp} lists their position
relative to the centre of M31, their total $F_{5007}$ flux of the
\Oiii\ emission with its corresponding detection limit and our best
estimate for their relative velocity $V_{\rm PN}$ and value of the
\OiiioHb\ ratio (or lower-limit when \Hb\ is not detected).
The latter two measurements are based on fits to spectra extracted
within a FWHM-wide aperture, in order to maximise the emission-line
signal and better isolate the kinematics and line ratio of PNe that in
projection are either close to each other or appear embedded in
diffuse gas.
These aperture spectra were also used to measure the stellar mean
velocity $V_{\star}$ and velocity dispersion $\sigma_{\star}$ along
the line-of-sight pointing to the detected PNe, which can be compared
to the PNe velocity $V_{\rm PN}$ in order to verify the PNe membership
to M31.
In fact, the presence of only 3 sources out of 12 with $| V_{\rm PN} -
V_{\star} | > \sigma_{\star}$ is consistent with the hypothesis that,
as a group, these PNe belong to M31.
Tab.~\ref{tab:PNeProp} also lists the values \OiiioHb\ ratio for our
detected sources, all of which, by construction, are above 3. Finally,
we note that these are the only PNe that we manage to isolate after
attempting to fit the most apparent peaks of \Oiii\ emission, and that
only in one case did our fit converge on an unresolved source with a
measured \OiiioHb\ $< 3$.

\placefigFOIIImapHST
\placefigFluxHSTcomparison
\subsection{Further Flux-Measurements Tests and Cross-check with HST data}
\label{subsec:AnalysisFurtherTests}

To test our purely spatial approach for isolating the PNe emission
from that of the diffuse gas, we re-measured the flux of three well
detected single PNe surrounded by extended emission (PNe 5, 6 and 8)
by means of a somewhat more general and simultaneous spectral and
spatial fit of the \Oiii\ emission.
More specifically, we re-fitted the 25 closer \sauron\ spectra to the
centre of each of these three PNe (in a $5\times5$ bins aperture) by
matching their \Oiii\ lines with two Gaussian components, one for the
PN emission and the other for the diffuse ionised-gas emission.
Across this region, we fixed the kinematics of each set of lines, and
still imposed a bi-dimensional Gaussian and constant spatial trend to
the flux of the Gaussian lines representing the PNe and diffuse-gas
components to the observed \Oiii\ profiles, respectively.
During the fit we then optimised the centre and amplitude of the PNe
profile, the level of the diffuse-gas emission, and also the position
and width of the PNe and diffuse-gas lines (even though the PNe
components were actually constrained to share a spectrally unresolved
profile).

The resulting values for the flux of the PNe matched in this way
differ by less than 15\% from our earlier estimates, which is a
tolerable level of discrepancy in the context of our analysis and
supports the use of a simpler and purely spatial approach for
extracting the flux also in the case of fainter PNe, where a spectral
decomposition is more uncertain.
Fig.~\ref{fig:FluxPNefits} also shows the best-fitting Gaussians and
constant radial profiles obtained for PNe 5, 6 and 8 through this more
refined fit (dashed lines), whereas Tab.~\ref{tab:PNeProp} lists the
corresponding $F_{5007}$ values.
These flux values do not change considerably if, for instance, the
background diffuse emission is treated as a plane rather than a
constant level in the framework of a purely spatial analysis or if,
alternatively, its amplitude is left free to vary across different
bins during the previously described spectral decomposition.

The presence in the Hubble Legacy Archive (HLA) of narrow-band Hubble
Space Telescope (\HST) images for the nuclear regions of M31 provides
a further opportunity to check whether we have missed any PN during
our source identification and, in the process, also to check again the
level of accuracy of our $F_{5007}$ estimates.
More specifically, we retrieved images taken with the Wide Field
Planetary Camera 2 (WFPC2) in the F502N, F606W and F547M filters and,
in this order, used them to a) initially identify possible sources of
\Oiii\ emission, b) confirm PNe candidates by detecting their \Ha\ and
\Nii\ emission and exclude spurious point-like sources in the F502N
band (due for instance to an imperfect treatment of cosmic rays), and
c) finally isolate stellar sources in the F502N images, most likely
bright red-giant branch stars (RGB). We then proceeded to measure the
$F_{5007}$ flux of the PNe thus identified in the F502N image through
standard aperture photometry.
%
%

Figs.~\ref{fig:FOIIImapHST} and \ref{fig:FluxHSTcomparison} show the
comparison between the positions and flux values of the PNe identified
in the \HST\ narrow-band images and through the spectral analysis of
the \sauron\ data.
Quite remarkably all the \sauron\ PNe sources in the field-of-view of
the F502N PC image do have an \HST\ counterpart (including our
marginally detected PN 12), which always falls within $\sim0\farcs1$
of the PN position estimated during our spatial fit to the
\sauron\ \Oiii\ flux values (e.g., Fig~\ref{fig:FluxPNefits}).
The converse is also true in Fig.~\ref{fig:FOIIImapHST}, except for PN
7 which (as already anticipated) was only recognized as part of an
apparent doublet with PN 3 following the analysis of the \HST\ images.
Fig.~\ref{fig:FluxHSTcomparison} shows that the $F_{5007}$ values
also agree quite well, generally within 30\% and with no strong
evidence for a systematic offset between the \HST\ and \sauron\ values
(Tab.~\ref{tab:PNeProp} also lists the $F_{5007}$ values derived from
the \HST\ images).
Our procedures for identifying PNe in the \sauron\ data and for
measuring their total \Oiii\ flux thus appear to be robust.

As a concluding aside, we note that our \HST\ analysis revealed also
the presence of two stellar sources, most likely RGB stars, in the
portion of the \sauron\ field-of-view covered by the WFPC2
observations (Fig.~\ref{fig:FOIIImapHST}). Incidentally, both stars
fall quite close to the brightest region of diffuse nebular emission
in our field, which is a coincidence that would be worth pursueing in
the scientific framework concerning the nature of the stellar sources
that are likely to power such diffuse gas \citep{Sar10}.

\section{Results}
\label{sec:Results}

In the context of understanding the connection between extragalactic
PNe and the properties of their parent stellar population, we now
consider the characteristics of the luminosity function of the PNe
that we have identified in the central $\sim$20 arcsec of M31, which
correspond to the inner $\sim 80$ pc at the adopted distance of 791
kpc to this galaxy (i.e. $m-M=24.49$ \citealp{Ton01}).

In particular, we wish to assess to what extent the central PNe of M31
could have been drawn from the generally adopted form of the PN
luminosity function (PNLF) introduced by \citet{Cia89}. This does not
necessarily have to be the case given that central stellar populations
of M31 are quite different, and in particular more metal-rich, than
what is observed in the rest of this galaxy (which is where the PNLF
form of \citeauthor{Cia89} was mostly established) and than what is
expected in the stellar halos of more distant galaxies where PNe are
typically detected.

\placefigPNLF
\placefigCompleteness

The central PNe luminosity function of M31 is presented in
Fig.~\ref{fig:PNLF}, where the $F_{5007}$ values of
Tab.~\ref{tab:PNeProp} have been first converted into apparent V-band
magnitudes following the $m_{5007} = -2.5\log{F_{5007}} -13.74$
formula of \cite{Cia89} and then into absolute $M_{5007}$ magnitudes
using a distance modulus of 24.49 magnitudes.
The solid line in Fig.~\ref{fig:PNLF} shows also the theoretical PNLF
of \citeauthor{Cia89}, which needs to be corrected for the fact that
progressively fainter PNe are harder to detect before testing whether
such a PNLF model can be regarded as the parent distribution for the
PNe that we observe.
Still, we already note a possible lack of bright PNe in M31 compared
to what would be expected from the standard form of the PNLF, which
cannot be explained by this kind of limitation.

To compute the completeness function of our experiment, that is, the
probability as a function of absolute magnitude $M_{5007}$ that a PN
of that brightness could be detected across the {\it entire\/}
field-of-view of the \sauron\ observations for M31, we used a set of
simulations similar to those performed in Paper~I in the case of M32,
even though here we also account for the adverse impact that the
presence of diffuse ionised-gas emission has on the detection of PNe.
These simulations, which are shown in Fig.~\ref{fig:Completeness} (see
also the figure caption for details), illustrate how PNe of an
absolute magnitude $M_{5007}\sim-1.0$ and fainter would quickly
escape detection over most of the \sauron\ field-of-view, which is
consistent with the fact that most of the detection limits for our PNe
sources (shown with open diamonds in Fig.~\ref{fig:PNLF}) pile up from
a $M_{5007}\sim-1.5$ and above.

The completeness-corrected standard form for the PNLF is shown by the
dashed line in Fig.~\ref{fig:PNLF}, whereas the grey dashed line
indicates the limit of our experiment if no diffuse ionised-gas
emission were present in the nuclear regions of M31, which allows a
comparison to our previous PNe survey in M32. In both cases, the
dotted lines show the associated uncertainties on these functions.
Accounting for incompleteness, a Kolgomorov-Smirnov test reveals that
there is just a 46\% probability that the observed PNe luminosity
distribution was drawn from the theoretical prescription of
\citeauthor{Cia89}, which is hardly surprising given the difference at
the high-luminosity end between such a model and our data.
In fact, a second set of simulations suggest that the apparent lack of
bright PNe in our data could only result from a rather unfortunate
sampling of the standard PNLF.
      
More specifically, starting from an intrinsic \citeauthor{Cia89} shape
of the PNLF and adopting a normalisation leading to match the number
of observed PNe once the PNLF is corrected for incompletess (as is in
fact the case for all the lines shown in Fig.~\ref{fig:PNLF}), we
generated a number of synthetic PN fields by considering at any
particular position in the \sauron\ field of view the probability of
having a PN of a given luminosity.
Such a probability function simply corresponds to the total intrinsic
PNLF rescaled by the fraction of stellar light that is observed in the
\sauron\ spatial bin that is being considered.
The \Oiii\ flux of the simulated PNe was then ``observed'' by
obtaining maps for the $A/rN$ ratio and by applying the same detection
criteria described in \S\ref{subsec:AnalysisPNeDetection}, including
checking whether a synthetic PN would be detected against the known
background of diffuse ionised-gas emission.
Based on a thousand of such synthetic PNe fields, only in 8.5\% of the
cases we could not find any PN brighter than $M_{5007}=-3.0$, the
current maximum PNe luminosity observed in the nuclear regions of M31.

This suggests that the standard form of the PNLF may not entirely
apply to the nuclear regions of M31, and that the observed dearth of
bright PNe in Fig.~\ref{fig:PNLF} may be real.

%
\section{Discussion}
\label{sec:Discussion}

The previous results on the luminosity function of the PNe found in
the very central regions of M31 (within 80 pc) are at odds with the
outcome of our earlier integral-field spectroscopic study of the PNe
population in the optical regions of M32 (within one $R_e$).
For that galaxy, the standard form of the PNLF could indeed be
considered, to a fair degree of confidence, as the parent distribution
of the PNe that we detected, with no significant features in their
observed distribution function.
We can attempt to interpret such a discrepancy by considering even the
differences between the properties of the stellar populations
encompassed by our \sauron\ observations for these two galaxies, in
particular as regards the average value of their metallicity.
A stronger stellar metallicity can indeed drive a larger mass-loss
rate efficiency during the RGB phase, which in turn would reduce the
number of HB stars with a massive H-shell that will eventually become
bright PNe.

Even though there exist numerous stellar population studies in the
literature on M31 and M32 (see, e.g., \citealp{Sag10} and
\citealp{Ros05}, respectively) we can use our own \sauron\ data to 
compare the mean values for the stellar age and metallicity in 
these two galaxies in a more consistent way than possible when using different absorption line-strength measurements from the literature.
For this, we co-added all the spectra in our data cubes after
subtracting the emission from both PNe and diffuse ionised-gas
components, and proceeded to fit such a total spectrum with the {\tt
  pPXF\/} method of \citet{Cap04}, using the entire MILES stellar
library of \citet{San06}.
We then combined the original stellar spectra from this library
according to the relative weights that they were assigned during the
{\tt pPXF\/} fit, in order to obtain what effectively is the optimal
stellar spectrum for representing the nuclear and central stellar
populations of M31 and M32, respectively.
Finally, we measured the strength of the standard \Hb, Fe5015 and
Mg$b$ indices from the Lick/IDS system on such optimal template, and
followed \citet{Kun10} to compute from the Fe5015 and Mg$b$ indices
the more purely metallicity-sensitive [MgbFe50]' index, which is
defined as $(0.69\times{\rm Mg}b+{\rm Fe5015})/2$.

The use of the MILES optimal templates allows us to circumvent
problems related to an imperfect sky subtraction and relative flux
calibration on the \sauron\ data, while obtaining at the same time
index values that are automatically corrected for kinematic
broadening.
Such an approach is recommended only if the templates can match almost
perfectly the real data, which is generally the case when using such a
large library as the MILES one, even when considering the spectra of
massive early-type galaxies \citep[e.g.,][]{Sar10,Oh11} where an
overabundance of $\alpha$ elements usually poses a problem for a
detailed spectral fitting (e.g., in the Mg$b$ region).

\placefigStellarPop

Fig.~\ref{fig:M31pop} shows the relative location of M31 and M32 in
the [MgFe50]' vs \Hb\ diagnostic diagram, where an estimate for the
mean age and metallicity of their nuclear and central stellar
populations, respectively, can be read through the model grids of
\citet{Sch07}.
In Fig.~\ref{fig:M31pop} we also plot the values for the same indices
for the early-type galaxies in the \sauron\ representative sample,
measured within both one $R_e$ and $R_e/8$ \citep[data from][]{Kun06},
and for the stellar halos of NGC~821 (at a distance of one $R_e$) and
NGC~3379 (at a distance between 3 and 4 $R_e$) from \citet{Wei09}.

The \sauron\ data confirm that the central 80 pc of M31 are more metal
rich than the stellar population found in the central regions of M32,
consistent with the hypothesis whereby an enhanced stellar-mass loss
in the RGB that is driven by a larger metallicity would eventually
lead to the observed lack of bright PNe in M31.
In fact, our PNe findings appear to be independently backed up by
\citet{Ros12} who, thanks to \HST\ observations, finds that the
relative number of bright post-AGB stars to that of fainter early
post-AGB or AGB-manqu\'e stars steadily decreases towards the centre
of M31, nicely following also the central metallicity gradient
measured by \citet{Sag10}.
These results would indeed agree with a dimming of the PNLF given that
only post-AGB stars can power the brightest PNe, while supporting also
the idea that a larger stellar metallicity is responsible for it.
As regards M32, on the other hand, Fig.~\ref{fig:M31pop} shows that
the optical regions of this galaxy have a value for the stellar
metallicity that is close, though not as low, to what \cite{Wei09}
estimated with \sauron\ data in the halos of the early-type galaxies
NGC~3379 and NGC~821.
Since also the measurements of \citet{Wei09} are based on \sauron\ 
data, all the values shown in Fig.~\ref{fig:M31pop} can be consistently 
compared to each other.
If metallicity is indeed related to the lack of post-AGB stars and
bright PNe in the nuclear regions of M31, then such a similarity may
help understand why the central PNe population of M32 appears on the
contrary to be consistent with the \citeauthor{Cia89}'s form for the
PNLF as in the case of the peripheral PNe populations of galaxies.

\citeauthor{Ros12} also show that the central metallicity gradient of
M31 corresponds closely to an increase in the far-UV excess towards
the centre of M31, much as observed when these quantities are measured
in different objects at larger galactic scales
\citep{Bur88,Bur11,Jeo12} and consistent with the presence of a large
number of AGB-manqu\'e stars. This contrasts again with the case of
M32, which has long been known to display little or no far-UV excess
flux \citep{Bur88}.

Finally, we note that in terms of their stellar metallicity the
nuclear regions of M31 fall roughly in between what is observed in the
central metal-rich stellar populations of early-type galaxies (within
$R_e/8$) and what is found on average within the effective radius
$R_e$ of these objects (Fig~\ref{fig:M31pop}, according always to
measurements based on \sauron\ data).
This suggests that also the PNe population of the optical regions of
early-type galaxies could display a similar dearth of bright PNe as
found in the central 80 pc of M31, which is a possibility that could
have a significant impact on the presently rather loose
anti-correlations between the specific number of PNe and either the
stellar metallicity or the strength of the UV-upturn
\citep{Buz06,Coc09}.
In fact, it may be even possible that the current trends reflect a
decrease only in the number of bright PNe in the halo of early-type
galaxies with progressively larger central (and presumably also
peripheral) metallicity values, rather than a change in the overall
specific content of PNe.
This may not be too far fatched considering that the completeness
limit of most PNe surveys in external galaxies does not extend beyond
1 to 2.5 magnitudes from the peak PN magnitude of $M_{5007}=-4.47$
(see, e.g., Tab.~5 in \citealp{Buz06}) so that we only have a limited
view on the shape of the PNLF of galactic halos.

\section{Conclusions}
\label{sec:Conclusions}

Building on our previous integral-field spectroscopic study for the
central PNe population of M32, in this paper we have further
investigated the very central $\sim$80 pc of M31 and found that:

\begin{itemize}

\item integral-field spectroscopy allows to detect PNe also in the
  presence of emission from diffuse gas, as commonly observed in
  early-type galaxies and in the bulge of spirals.

\item our relative inexpensive \sauron\ data could spot nearly all the
  PNe found in \HST\ narrow-band band images, measuring total
  \Oiii$\lambda5007$ flux values well in agrement with the
  \HST\ measurements.

\item contrary to the case of the central regions of M32, the nuclear
  PNe population of M31 is only marginally consistent with the
  generally adopted \citeauthor{Cia89} form of the PNe luminosity
  function

\item such a discrepancy is due to an observed lack of PNe with absolute
  magnitude $M_{5007}$ brighter than $-3.0$, which simulations suggest would occur only considering a 
  rather unfortunate (in 8.5\% of the cases) sampling of such a model PNLF.

\end{itemize}

Considering that the nuclear stellar population of M31 is
characterised by a larger stellar metallicity and a much stronger
far-UV excess compared to what is found in the central (within one
effective radius) regions of M32, the previous results would appear to
support the idea that a larger metallicity (which enhances the
stellar-mass loss efficiency in the RGB) can lead to an
horizontal-branch population that is more tilted toward less massive
and hotter He-burning stars, so that its progeny consists mostly of
UV-bright AGB-manqu\'e stars, but few, if any, bright PNe.

A lack of bright PNe is also consistent with the recent reports by
\citet{Ros12} on a shortage of post-AGB stars towards the nucleus of
M31, since bright PNe are powered by central post-AGB stars.
If such a dearth of post-AGB stars and bright PNe does indeed
correspond to nearly-solar values for the stellar metallicity or
above, then such a feature is likely to characterise also the PNe
luminosity function in the optical regions of more distant and massive
old stellar systems, 
which would imply that the PNLF form of
\citet{Cia89} is not universally applicable.
This is in fact a possibility that will be explored in future papers
that will use the \sauron\ data already at hand for the complete
ATLAS$^{\rm 3D}$ sample of early-type galaxies \citep{Cap11},
circumnventing the need for a prohibitively expensive narrow-band
imaging campaign with \HST.

\section*{Acknowledgements}
We are grateful to Laura Greggio, Alvio Renzini, Enrico Maria Corsini, 
and Harald Kuntschner for their suggestions. 
We also thank the referee Michael G. Richer for providing very useful comments 
that improved the quality of this manuscript. 
NP acknowledges the University of Hertfordshire for funding his visit at 
the Centre for Astrophysics research where most of this study was carried out. 
NP and MS are also grateful, respectively, to the Astronomy Department of the 
University of Padova and to the European Southern Observatory for their hospitality. 
Finally, MS and MC acknowledge their respective STFC Advanced 
(ST/F009186/1) and Royal Society University Research fellowships.

%

\label{lastpage}
\end{document}